\documentclass[12pt]{article}
\usepackage{epsfig}

\begin{document}

\begin{titlepage}

\begin{flushright}
CLNS~04/1888\\
{\tt hep-ph/0408208}\\[0.2cm]
August 18, 2004
\end{flushright}

\vspace{0.7cm}
\begin{center}
\Large\bf\boldmath 
Soft-Collinear Factorization and the Calculation of the $B\to X_s\gamma$ 
Rate\unboldmath
\end{center}

\vspace{0.8cm}
\begin{center}
{\sc Matthias Neubert\footnote{Invited talk presented at the {\em 39th 
Rencontres de Moriond: Electroweak Interactions and Unified Theories\/} (La 
Thuile, Italy, 21--28 March 2004), and at {\em Continuous Advances in QCD 
2004\/} (Minneapolis, Minnesota, 13--16 May 2004)}}\\
\vspace{0.7cm}
{\sl Institute for High-Energy Phenomenology\\
Newman Laboratory for Elementary-Particle Physics, Cornell University\\
Ithaca, NY 14853, U.S.A.}
\end{center}

\vspace{1.0cm}
\begin{abstract}
\vspace{0.2cm}\noindent
Using results on soft-collinear factorization for inclusive $B$-meson decay 
distributions, a systematic study of the partial $B\to X_s\gamma$ decay rate 
with a cut $E_\gamma\ge E_0$ on photon energy is performed. For values of 
$E_0\le 1.9$\,GeV the rate can be calculated without reference to shape 
functions. The result depends on three large scales: $m_b$, 
$\sqrt{m_b\Delta}$, and $\Delta=m_b-2E_0$. The sensitivity to the scale 
$\Delta\approx 1.1$\,GeV (for $E_0\approx 1.8$\,GeV) introduces significant 
uncertainties, which have been ignored in the past. Our new prediction for the 
$B\to X_s\gamma$ branching ratio with $E_\gamma\ge 1.8$\,GeV is 
$\mbox{Br}(B\to X_s\gamma)=(3.44\pm 0.53\pm 0.35)\times 10^{-4}$, where the 
errors refer to perturbative and parameter uncertainties, respectively. The 
implications of larger theory uncertainties for New Physics searches are 
explored with the example of the type-II two-Higgs-doublet model.
\end{abstract}

\end{titlepage}

\section{Introduction}

Given the prominent role of $B\to X_s\gamma$ decay in searching for physics 
beyond the Standard Model, it is of great importance to have a precise 
prediction for its inclusive rate and CP asymmetry in the Standard Model. The 
total inclusive $B\to X_s\gamma$ decay rate can be calculated using a 
conventional operator-product expansion (OPE) based on an expansion in 
logarithms and inverse powers of the $b$-quark mass. However, in practice 
experiments can only measure the high-energy part of the photon spectrum, 
$E_\gamma\ge E_0$, where typically $E_0=2$\,GeV or slightly below (measured in 
the $B$-meson rest frame) \cite{Chen:2001fj,Koppenburg:2004fz}. With 
$E_\gamma$ restricted to be close to the kinematic endpoint at $M_B/2$, the 
hadronic final state $X_s$ is constrained to have large energy $E_X\sim M_B$ 
but only moderate invariant mass $M_X\sim(M_B\Lambda_{\rm QCD})^{1/2}$. In 
this kinematic region, an infinite number of leading-twist terms in the OPE 
need to be resummed into a non-perturbative shape function, which describes 
the momentum distribution of the $b$-quark inside the $B$ meson 
\cite{Neubert:1993ch,Bigi:1993ex}.

Conventional wisdom based on phenomenological studies of shape-function 
effects says these effects are important near the endpoint of the photon 
spectrum, but they can be ignored as soon as the cutoff $E_0$ is lowered below 
about 1.9\,GeV. In other words, there should be an instantaneous transition 
from the ``shape-function region'' of large non-perturbative corrections to 
the ``OPE region'', in which hadronic corrections to the rate are suppressed 
by at least two powers of $\Lambda_{\rm QCD}/m_b$. Below, we argue that this 
notion is based on a misconception. While it is correct that once the cutoff 
$E_0$ is chosen below 1.9\,GeV the decay rate can be calculated using a local 
short-distance expansion, we show that this expansion involves three ``large'' 
scales. In addition to the hard scale $m_b$, an intermediate scale 
$\sqrt{m_b\Delta}$ corresponding to the typical invariant mass of the hadronic 
final state $X_s$, and a low scale $\Delta=m_b-2E_0$ related to the width of 
the energy window over which the measurement is performed, become of crucial 
importance. The precision of the theoretical calculations is ultimately 
determined by the value of the lowest short-distance scale $\Delta$, which in 
practice is of order 1\,GeV or only slightly larger. The theoretical accuracy 
that can be reached is therefore not as good as in the case of a conventional 
heavy-quark expansion applied to the $B$ system. More likely, it is similar to 
(if not worse than) the accuracy reached, say, in the description of the 
inclusive hadronic decay rate of the $\tau$ lepton.

While we are aware that this conclusion may come as a surprise to many 
practitioners in the field of flavor physics, we believe that it is an 
unavoidable consequence of our analysis. Not surprisingly, then, we find that 
the error estimates for the $B\to X_s\gamma$ branching ratio that can be found 
in the literature are, without exception, too optimistic. Since there are 
unknown $\alpha_s^2(\Delta)$ corrections at the low scale $\Delta\sim 1$\,GeV, 
we estimate the present perturbative uncertainty in the $B\to X_s\gamma$ 
branching ratio with $E_0$ in the range between 1.6 and 1.8\,GeV to be of 
order 10--15\%. In addition, there are uncertainties due to other sources, 
such as the $b$- and $c$-quark masses. The combined theoretical uncertainty is 
of order 15--20\%, about twice as large as what has been claimed in the past. 
While this is a rather pessimistic conclusion, we stress that the uncertainty 
is limited by unknown, higher-order perturbative terms, not by 
non-perturbative effects, which we find to be under good control. Therefore, 
there is room for a reduction of the error by means of well-controlled 
perturbative calculations.

\section{QCD factorization theorem}

Using recent results on the factorization of inclusive $B$-meson decay 
distributions \cite{Bosch:2004th,Bauer:2003pi}, it is possible to derive a QCD 
factorization formula for the integrated $B\to X_s\gamma$ decay rate with a 
cut $E_\gamma\ge E_0$ on photon energy. In the region of large $E_0$, the 
leading contribution to the rate can be factorized in the form \cite{mywork}
\begin{eqnarray}\label{ff}
   \Gamma_{\bar B\to X_s\gamma}^{\rm leading}(E_0)
   &=& \frac{G_F^2\alpha}{32\pi^4}\,|V_{tb} V_{ts}^*|^2\,
    \overline{m}_b^2(\mu_h)\,|H_\gamma(\mu_h)|^2\,U_1(\mu_h,\mu_i) \\
   &\times& \int_0^{\Delta_E}\!dP_+\,(M_B-P_+)^3
    \int_0^{P_+}\!d\hat\omega\,m_b\,J\big(m_b(P_+-\hat\omega),\mu_i\big)\,
    \hat S(\hat\omega,\mu_i) \,, \nonumber
\end{eqnarray}
where $\Delta_E=M_B-2E_0$ is twice the width of the window in photon energy 
over which the measurement of the decay rate is performed. The variable 
$P_+=E_X-|\vec{P}_X|$ is the ``plus component'' of the 4-momentum of the 
hadronic final state $X_s$, which is related to the photon energy by 
$P_+=M_B-2E_\gamma$. The endpoint region of the photon spectrum is defined by 
the requirement that $P_+\le\Delta_E\ll M_B$, in which case $P^\mu$ is called 
a hard-collinear momentum \cite{Bosch:2003fc}.

In the factorization formula, $\mu_h\sim m_b$ is a hard scale, while 
$\mu_i\sim\sqrt{m_b\Lambda_{\rm QCD}}$ is an intermediate hard-collinear scale 
of order the invariant mass of the hadronic final state. The precise values of 
these matching scales are irrelevant, since the rate is formally independent 
of $\mu_h$ and $\mu_i$. The hard corrections captured by the function 
$H_\gamma(\mu_h)$ result from the matching of the effective weak Hamiltonian 
of the Standard Model (or any of its extensions) onto a leading-order current 
operator of soft-collinear effective theory (SCET) \cite{Bauer:2000yr}. At 
tree level, $H_\gamma(\mu_h)=C_{7\gamma}^{\rm eff}(\mu_h)$ is equal to the 
``effective'' coefficient 
$C_{7\gamma}^{\rm eff}=C_{7\gamma}-\frac13\,C_5-C_6$. The expression valid at 
next-to-leading order can be found in Ref.~\cite{mywork}. The function 
$H_\gamma(\mu_h)$ is multiplied by the running $b$-quark mass 
$\overline{m}_b(\mu_h)$ defined in the $\overline{\rm MS}$ scheme, which is 
part of the electromagnetic dipole operator $Q_{7\gamma}$. 

The jet function $J\big(m_b(P_+-\hat\omega),\mu_i\big)$ in (\ref{ff}) 
describes the physics of the final-state hadronic jet. An expression for this
function valid at next-to-leading order in perturbation theory has been 
derived in Refs.~\cite{Bosch:2004th,Bauer:2003pi}. The perturbative expansion 
of the jet function can be trusted as long as $\mu_i^2\sim m_b\Delta$ with 
$\Delta=m_b-2E_0\ll M_B$. Note that the ``natural'' choices $\mu_h\propto m_b$ 
and $\mu_i^2\equiv m_b\,\tilde\mu_i$ with $\tilde\mu_i$ independent of $m_b$ 
remove all reference to the $b$-quark mass (other than in the arguments of 
running coupling constants) from the factorization formula.

The shape function $\hat S(\hat\omega,\mu_i)$ parameterizes our ignorance 
about the soft physics associated with bound-state effects inside the $B$ 
meson \cite{Neubert:1993ch,Bigi:1993ex}. Its naive interpretation is that of a 
parton distribution function, governing the distribution of the light-cone 
component $k_+$ of the residual momentum of the $b$ quark inside the heavy 
meson. Once radiative corrections are included, however, a probabilistic 
interpretation of the shape function breaks down \cite{Bosch:2004th}. For 
convenience, the shape function is renormalized in (\ref{ff}) at the 
intermediate hard-collinear scale $\mu_i$ rather than at a hadronic scale 
$\mu_{\rm had}$. This removes any uncertainties related to the evolution from 
$\mu_i$ to $\mu_{\rm had}$. Since the shape function is universal, all that 
matters is that it is renormalized at the same scale when comparing different 
processes.

The last ingredient in the factorization formula is the function 
$U_1(\mu_h,\mu_i)$, which describes the renormalization-group (RG) evolution 
of the hard function $|H_\gamma|^2$ from the high matching scale $\mu_h$ down 
to the intermediate scale $\mu_i$, at which the jet and shape functions are 
renormalized. The exact expression for this quantity and its perturbative
expansion valid at next-to-next-to-leading logarithmic order can be found in 
Ref.~\cite{mywork}.

\begin{figure}
\begin{center}
\epsfig{file=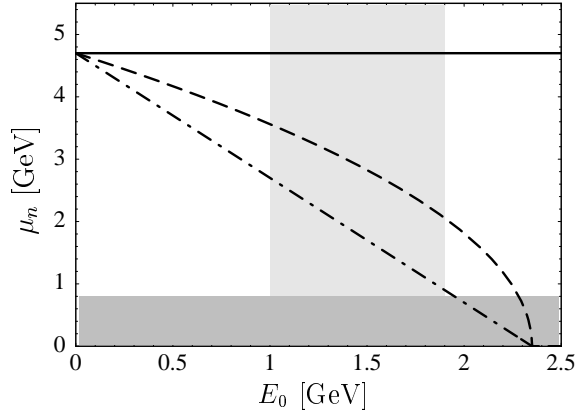,width=7.5cm}
\end{center}
\vspace{-0.4cm}
\caption{\label{fig:scales}
Dependence of the three scales $\mu_h=m_b$ (solid), $\mu_i=\sqrt{m_b\Delta}$ 
(dashed), and $\mu_0=\Delta$ (dash-dotted) on the cutoff $E_0$, assuming 
$m_b=4.7$\,GeV. The gray area at the bottom shows the domain of 
non-perturbative physics. The light gray band in the center indicates the 
region where the MSOPE must be applied.}
\end{figure}

As written in (\ref{ff}), the decay rate is sensitive to non-perturbative
hadronic physics via its dependence on the shape function. This sensitivity is 
unavoidable as long as the scale $\Delta=m_b-2E_0$ is a hadronic scale,
corresponding to the endpoint region of the photon spectrum above, say, 
2\,GeV. Here we are interested in a situation where $E_0$ is lowered out of 
the shape-function region, such that $\Delta$ can be considered large compared 
with $\Lambda_{\rm QCD}$. For orientation, we note that with $m_b=4.7$\,GeV 
and the cutoff $E_0=1.8$\,GeV employed in a recent analysis by the Belle 
Collaboration \cite{Koppenburg:2004fz} one gets $\Delta=1.1$\,GeV. The values 
of the three relevant physical scales as functions of the photon-energy 
cutoff $E_0$ are shown in Figure~\ref{fig:scales}. This plot illustrates the 
fact that the transition from the shape-function region to the region where a 
conventional OPE can be applied is not abrupt but proceeds via an intermediate 
region, in which a short-distance analysis based on a multi-stage OPE (MSOPE) 
can be performed. The transition from the shape-function region into the MSOPE 
region occurs when the scale $\Delta$ becomes numerically (but not 
parametrically) large compared with $\Lambda_{\rm QCD}$. Then terms of order 
$\alpha_s^n(\Delta)$ and $(\Lambda_{\rm QCD}/\Delta)^n$, which are 
non-perturbative in the shape-function region, gradually become decent 
expansion parameters. Only for very low values of the cutoff ($E_0<1$\,GeV or 
so) it is justified to treat $\Delta$ and $\sqrt{m_b\Delta}$ as scales of 
order $m_b$. 

Separating the contributions associated with these scales requires a 
sophisticated multi-step procedure. The first step, the separation of the hard 
scale from the intermediate scale, has already been achieved in (\ref{ff}). To 
proceed further, we use that integrals of smooth weight functions with the 
shape function $\hat S(\hat\omega,\mu)$ can be expanded in a series of forward 
$B$-meson matrix elements of local operators in heavy-quark effective 
theory (HQET) \cite{Neubert:1993mb}, provided that the integration domain is 
large compared with $\Lambda_{\rm QCD}$ \cite{Bosch:2004th,Bauer:2003pi}. The 
perturbative expansions of the associated Wilson coefficient functions can be 
trusted as long as $\mu\sim\Delta$. In order to complete the scale separation, 
it is therefore necessary to evolve the shape function in (\ref{ff}) from the 
intermediate scale $\mu_i\sim\sqrt{m_b\Delta}$ down to a scale 
$\mu_0\sim\Delta$. This can be achieved using the analytic solution to the 
integro-differential RG evolution equation for the shape function in momentum 
space \cite{Bosch:2004th,Lange:2003ff}.

As a final comment, we stress that the main purpose of performing the scale
separation using the MSOPE is not that this allows us to resum Sudakov 
logarithms. Indeed, the ``large logarithm'' $\ln(m_b/\Delta)\approx 1.5$ is 
only parametrically large, but not numerically. What is really important is to 
disentangle the physics at the low scale $\mu_0\sim\Delta$, which is ``barely 
perturbative'', from the physics associated with higher scales, where a 
short-distance treatment is on much safer grounds. The MSOPE allows us to 
distinguish between the three coupling constants $\alpha_s(m_b)\approx 0.22$, 
$\alpha_s(\sqrt{m_b\Delta})\approx 0.29$, and $\alpha_s(\Delta)\approx 0.44$
(for $\Delta=1.1$\,GeV), which are rather different despite the fact that 
there are no numerically large logarithms in the problem. Given the values of 
these couplings, we expect that scale separation between $\Delta$ and $m_b$ is 
as important as that between $m_b$ and the weak scale $M_W$.

\section{Calculation of the shape-function integral}

The scale dependence of the renormalized shape function is governed by an 
integro-differential RG evolution equation, whose exact solution in momentum 
space can be found using a technique developed in Ref.~\cite{Lange:2003ff}. 
The result takes the remarkably simple form
\begin{equation}\label{sonice}
   \hat S(\hat\omega,\mu_i) = U_2(\mu_i,\mu_0)\,
   \frac{e^{-\gamma_E\eta}}{\Gamma(\eta)} \int_0^{\hat\omega}\!d\hat\omega'\,
   \frac{\hat S(\hat\omega',\mu_0)}
        {\mu_0^\eta(\hat\omega-\hat\omega')^{1-\eta}} \,.
\end{equation}
The exact expression for the evolution function $U_2(\mu_i,\mu_0)$ can be 
found in Ref.~\cite{mywork}, and  
\begin{equation}
   \eta = \int\limits_{\alpha_s(\mu_i)}^{\alpha_s(\mu_0)}\!
   d\alpha\,\frac{2\Gamma_{\rm cusp}(\alpha)}{\beta(\alpha)}
   = \frac{\Gamma_0}{\beta_0}\ln\frac{\alpha_s(\mu_0)}{\alpha_s(\mu_i)}
   + \dots
\end{equation}
is given in terms of the cusp anomalous dimension \cite{Korchemsky:wg}.

Relation (\ref{sonice}) accomplishes the evolution of the shape function from 
the intermediate scale down to the low scale $\mu_0\sim\Delta$. The remaining 
task is to expand the integral over the shape function in (\ref{ff}) in a 
series of forward $B$-meson matrix elements of local HQET operators of 
increasing dimension, multiplied by perturbative coefficient functions. This 
can be done whenever $\Delta=\Delta_E-\bar\Lambda=m_b-2E_0$ is large compared 
with $\Lambda_{\rm QCD}$ \cite{Bosch:2004th}. The result is
\begin{eqnarray}\label{rate2}
   \Gamma_{\bar B\to X_s\gamma}^{\rm leading}(E_0)
   &=& \frac{G_F^2\alpha}{32\pi^4}\,|V_{tb} V_{ts}^*|^2\,m_b^3\,
    \overline{m}_b^2(\mu_h)\,|H_\gamma(\mu_h)|^2\,U_1(\mu_h,\mu_i)\,
    U_2(\mu_i,\mu_0) \nonumber\\
   &\times& \frac{e^{-\gamma_E\eta}}{\Gamma(1+\eta)}
    \left( \frac{\Delta}{\mu_0} \right)^\eta \left[
    1 + \frac{C_F\alpha_s(\mu_i)}{4\pi}\,{\cal J}(\Delta)
    + \frac{C_F\alpha_s(\mu_0)}{4\pi}\,{\cal S}(\Delta) \right]
    \nonumber\\
   &\times&  \left[ 1 + \frac{\eta(\eta-1)}{2}\,
    \frac{(-\lambda_1)}{3\Delta^2} + \dots \right] ,
\end{eqnarray}
where 
\begin{eqnarray}
   {\cal J}(\Delta)
   &=& 2 \ln^2\!\frac{m_b\Delta}{\mu_i^2}
    - \big[ 4h(\eta) + 3 \big] \ln\frac{m_b\Delta}{\mu_i^2}
    + 2h^2(\eta) + 3h(\eta) - 2h'(\eta) + 7 - \frac{2\pi^2}{3} \,, \nonumber\\
   {\cal S}(\Delta)
   &=& - 4 \ln^2\!\frac{\Delta}{\mu_0}
   + 4 \left[ 2h(\eta) - 1 \right] \ln\frac{\Delta}{\mu_0}
   - 4h^2(\eta) + 4h(\eta) + 4h'(\eta) - \frac{5\pi^2}{6} \,,
\end{eqnarray}
and $h(\eta)=\psi(1+\eta)+\gamma_E$ is the harmonic function generalized to 
non-integer argument. Even though it is parametrically larger than ordinary 
power corrections of order $(\Lambda_{\rm QCD}/m_b)^2$, the ``enhanced'' 
$\lambda_1/\Delta^2$ correction in (\ref{rate2}) remains small in the region 
of ``perturbative'' values of $\Delta$, where the MSOPE can be trusted. The 
net effect amounts to a reduction of the decay rate by less than 5\%.

The rate in (\ref{rate2}) is formally independent of the three matching 
scales, at which we switch from QCD to SCET ($\mu_h$), from SCET to HQET 
($\mu_i$), and finally at which the shape-function integral is expanded in a 
series of local operators ($\mu_0$). In practice, a residual scale dependence 
arises because we have truncated the perturbative expansion. Varying the three 
matching scales about their default values provides some information about 
unknown higher-order terms. In the limit where the intermediate and low 
matching scales $\mu_i$ and $\mu_0$ are set equal to the hard matching scale 
$\mu_h$, our result reduces to the conventional formula used in previous 
analyses of the $B\to X_s\gamma$ decay rate. However, this choice cannot be 
justified on physical grounds.

In (\ref{rate2}) we have accomplished a complete resummation of 
(parametrically) large logarithms at next-to-next-to-leading logarithmic order 
in RG-improved perturbation theory, which is necessary in order to calculate 
the decay rate with ${\cal O}(\alpha_s)$ accuracy. Specifically, it means that 
terms of the form $\alpha_s^n L^k$ with $k=(n-1),\dots,2n$ and 
$L=\ln(m_b/\Delta)$ are correctly resummed to all orders in perturbation 
theory. To the best of our knowledge, a complete resummation at 
next-to-next-to-leading order has never been achieved before. Finally, we 
stress that the various next-to-leading order terms in the expression for the 
decay rate should be consistently expanded to order $\alpha_s$ before applying 
our results to phenomenology.

Up to this point, the $b$-quark mass $m_b$ entering the formula (\ref{rate2})
for the decay rate is defined in the on-shell scheme. While this is most 
convenient for performing calculations using heavy-quark expansions, it is 
well known that HQET parameters defined in the pole scheme suffer from 
infra-red renormalon ambiguities. It is necessary to replace them in favor of 
some physical, short-distance parameters. For our purposes, the 
``shape-function scheme'' provides for a particularly suitable definition of 
the heavy-quark mass \cite{Bosch:2004th}. The idea is that a good estimate of 
a shape-function integral can be obtained using the mean-value theorem, 
replacing $\hat\omega$ with
\begin{equation}
   \langle\hat\omega\rangle_\Delta =
   \frac{\int_0^{\Delta_E}\!d\hat\omega\,\hat\omega\,\hat S(\hat\omega,\mu_0)}
        {\int_0^{\Delta_E}\!d\hat\omega\,\hat S(\hat\omega,\mu_0)}
   \equiv \bar\Lambda(\Delta,\mu_0) = M_B - m_b(\Delta,\mu_0) \,.
\end{equation}
Here $m_b(\Delta,\mu_0)$ is the running shape-function mass, which depends on 
a hard cutoff $\Delta$ in addition to the renormalization scale $\mu_0$. The 
quantity $\Delta$ in the shape-function scheme is defined by the implicit 
equation $\Delta=\Delta_E-\bar\Lambda(\Delta,\mu_0)=m_b(\Delta,\mu_0)-2E_0$. 
The shape-function scheme provides a physical definition of $m_b$, which can 
be related to any other short-distance definition using perturbation theory. 
Based on various sources of phenomenological information including 
$\Upsilon$ spectroscopy and moments of inclusive $B$-meson decay spectra, the
value of the shape-function mass at a reference scale $\mu_*=1.5$\,GeV has 
been determined as $m_b(\mu_*,\mu_*)=(4.65\pm 0.07)$\,GeV \cite{Bosch:2004th}.

The results discussed so far provide a complete description of the 
$B\to X_s\gamma$ decay rate at leading order in the $1/m_b$ expansion, where 
the two-step matching QCD\,$\to$\,SCET\,$\to$\,HQET is well understood. For 
practical applications, however, it is necessary to include corrections 
arising at higher orders in the heavy-quark expansion. Most important are 
``kinematic'' power corrections of order $(\Delta/m_b)^n$, which are not 
associated with new hadronic parameters. Unlike the non-perturbative 
corrections, these effects appear already at first order in $\Delta/m_b$, and 
they are numerically dominant in the region where 
$\Delta\gg\Lambda_{\rm QCD}$. Technically, the kinematic power corrections 
correspond to subleading jet functions arising in the matching of QCD onto 
higher-dimensional SCET operators, as well as subleading shape functions 
arising in the matching of SCET onto HQET operators. The corresponding terms 
are known in fixed-order perturbation theory, without scale separation and RG 
resummation \cite{Greub:1996tg,Kagan:1998ym}. To perform a complete RG 
analysis of even the first-order terms in $\Delta/m_b$ is beyond the scope of 
our discussion. Since for typical values of $E_0$ the power corrections only 
account for about 15\% of the $B\to X_s\gamma$ decay rate, an approximate 
treatment suffices at the present level of precision. Details of how these 
corrections are implemented can be found in Ref.~\cite{mywork}.

\section{Ratios of decay rates}

The contributions from the three different short-distance scales entering the 
central result (\ref{rate2}) and the associated theoretical uncertainties can 
be disentangled by taking ratios of decay rates. Some ratios probe truly
short-distance physics (i.e., physics above the scale $\mu_h\sim m_b$) and so 
remain unaffected by the new theoretical results presented above. For some 
other ratios, the short-distance physics associated with the hard scale 
cancels to a large extent, so that one probes physics at the intermediate and 
low scales, irrespective of the short-distance structure of the theory. 

\paragraph{Ratios insensitive to low-scale physics:}
Physics beyond the Standard Model may affect the theoretical results for the 
$B\to X_s\gamma$ branching ratio and CP asymmetry only via the Wilson 
coefficients of the various operators in the effective weak Hamiltonian. As a 
result, the ratio of the $B\to X_s\gamma$ decay rate in a New-Physics model 
relative to that in the Standard Model remains largely unaffected by the 
resummation effects studied in the present work. From (\ref{rate2}), we obtain
\begin{equation}
   \frac{\Gamma_{\bar B\to X_s\gamma}|_{\rm NP}}
        {\Gamma_{\bar B\to X_s\gamma}|_{\rm SM}}
   = \frac{|H_\gamma(\mu_h)|_{\rm NP}^2}{|H_\gamma(\mu_h)|_{\rm SM}^2}
   + \mbox{power corrections.}
\end{equation}
The power corrections would introduce some mild dependence on the intermediate 
and low scales $\mu_i$ and $\mu_0$, as well as on the cutoff $E_0$.

Another important example is the direct CP asymmetry in $B\to X_s\gamma$ 
decays, for which we obtain
\begin{equation}
   A_{\rm CP}
   = \frac{\Gamma_{\bar B\to X_s\gamma} - \Gamma_{B\to X_{\bar s}\gamma}}
          {\Gamma_{\bar B\to X_s\gamma} + \Gamma_{B\to X_{\bar s}\gamma}}
   = \frac{|H_\gamma(\mu_h)|^2 - |\overline{H}_\gamma(\mu_h)|^2}
          {|H_\gamma(\mu_h)|^2 + |\overline{H}_\gamma(\mu_h)|^2}
   + \mbox{power corrections,}
\end{equation}
where $\overline{H}_\gamma$ is obtained from $H_\gamma$ by CP conjugation, 
which in the Standard Model amounts to replacing the CKM matrix elements by 
their complex conjugates. It follows that predictions for the CP asymmetry in 
the Standard Model and various New Physics scenarios \cite{Kagan:1998bh} 
remain largely unaffected by our considerations.

\paragraph{Ratios sensitive to low-scale physics:}
The multi-scale effects studied in this work result from the fact that in 
practice the $B\to X_s\gamma$ decay rate is measured with a restrictive cut on 
the photon energy. These complications would be absent if it were possible to 
measure the fully inclusive rate. It is convenient to define a function 
$F(E_0)$ as the ratio of the $B\to X_s\gamma$ decay rate with a cut $E_0$ 
divided by the total rate, 
\begin{equation}
   F(E_0) = \frac{\Gamma_{\bar B\to X_s\gamma}(E_0)}
                 {\Gamma_{\bar B\to X_s\gamma}(E_*)} \,. 
\end{equation}
Because of a logarithmic soft-photon divergence for very low energy, it is 
conventional \cite{Kagan:1998ym} to define the ``total'' inclusive rate as the 
rate with a very low cutoff $E_*=m_b/20$. The denominator in the expression 
for $F(E_0)$ can be evaluated using the standard OPE, which corresponds to 
setting all three matching scales equal to $\mu_h$. The numerator is given by 
our expression in (\ref{rate2}), supplemented by power corrections.

Another important example of a ratio that is largely insensitive to the hard 
matching contributions is the average photon energy $\langle E_\gamma\rangle$, 
which has been proposed as a good way to measure the $b$-quark 
mass \cite{Kapustin:1995nr}. The impact of shape-function effects on the 
theoretical prediction for this quantity has been investigated and was found 
to be significant \cite{Kagan:1998ym,Bigi:2002qq}. Here we study the average 
photon energy in the MSOPE region, where a model-independent prediction can be 
obtained. It is structurally different from the one found using the 
conventional OPE in the sense that contributions associated with different 
scales are disentangled from each other. We stress that the hard scale 
$\mu_h\sim m_b$ affects the average photon energy only via second-order power 
corrections. This shows that it is not appropriate to compute the quantity 
$\langle E_\gamma\rangle$ using a simple heavy-quark expansion at the scale 
$m_b$, which is however done in the conventional 
approach \cite{Kapustin:1995nr}. This observation is important, because 
information about moments of the $B\to X_s\gamma$ photon spectrum is sometimes 
used in global fits to determine the CKM matrix element $|V_{cb}|$. Keeping 
only the leading power corrections, which is a very good approximation, we 
find that $\langle E_\gamma\rangle$ only depends on physics at the 
intermediate and low scales $\mu_i$ and $\mu_0$. For $E_0=1.8$\,GeV, we obtain 
$\langle E_\gamma\rangle\approx[2.222+0.254\alpha_s(\sqrt{m_b\Delta})%
+0.009\alpha_s(\Delta)]\,\mbox{GeV}\approx 2.30$\,GeV.

\section{Numerical results}

We are now ready to present the phenomenological implications of our findings. 
A complete list of the relevant input parameters and their uncertainties is 
given in Ref.~\cite{mywork}, where we also explain our strategy for estimating 
the perturbative uncertainty as well as the uncertainty due to parameter 
variations.

We begin by presenting predictions for the CP-averaged $B\to X_s\gamma$ 
branching fraction with a cutoff $E_\gamma\ge E_0$ applied on the photon 
energy measured in the $B$-meson rest frame. Lowering $E_0$ below 2\,GeV is 
challenging experimentally. The first measurement with $E_0=1.8$\,GeV has 
recently been reported by the Belle Collaboration \cite{Koppenburg:2004fz}. It 
yields\footnote{To obtain the first result we had to undo a theoretical 
correction accounting for the effects of the cut $E_\gamma>1.8$\,GeV, which 
had been applied to the experimental data.}
\begin{eqnarray}\label{belle}
   \mbox{Br}(B\to X_s\gamma) \Big|_{E_\gamma>1.8\,{\rm GeV}}
   &=& (3.38\pm 0.30\pm 0.29)\cdot 10^{-4} \,, \nonumber\\
   \langle E_\gamma\rangle \Big|_{E_0=1.8\,{\rm GeV}}
   &=& (2.292\pm 0.026\pm 0.034)\,\mbox{GeV} \,.
\end{eqnarray}
For $E_0=1.8$\,GeV we have $\Delta\approx 1.1$\,GeV, which is sufficiently 
large to apply the formalism developed in the present work. (For comparison, 
the value $E_0=2.0$\,GeV adopted in the CLEO analysis \cite{Chen:2001fj} 
implies $\Delta\approx 0.7$\,GeV, which we believe is too low for a 
short-distance treatment.) We find
\begin{equation}
   \mbox{Br}(B\to X_s\gamma) \Big|_{E_0=1.8\,{\rm GeV}}
   = (3.44\pm 0.53\,[\mbox{pert.}]\pm 0.35\,[\mbox{pars.}])\times 10^{-4} \,,
\end{equation}
where the first error refers to the perturbative uncertainty and the second 
one to parameter variations. The largest parameter uncertainties are due to 
the $b$- and $c$-quark masses. Our result is in excellent agreement with the 
experimental value shown in (\ref{belle}). Comparing the two results, and 
naively assuming Gaussian errors, we conclude that
\begin{equation}
   \mbox{Br}(B\to X_s\gamma)_{\rm exp} - \mbox{Br}(B\to X_s\gamma)_{\rm SM}
   < 1.4\cdot 10^{-4} \quad \mbox{(95\% CL)} \,.
 \end{equation}
Mainly as a result of the enlarged theoretical uncertainty, this bound is much 
weaker than the one derived in Ref.~\cite{Gambino:2001ew}, where this 
difference was found to be less than $0.5\cdot 10^{-4}$. Hence, we obtain a 
much weaker constraint on New Physics parameters. For instance, for the case 
of the type-II two-Higgs-doublet model, we may use the analysis of 
Ref.~\cite{Borzumati:1998tg} to deduce
\begin{equation}
   m_{H^+} > \mbox{(slightly below) 200\,GeV} \quad \mbox{(95\% CL)} \,,
\end{equation}
which is significantly weaker than the constraints $m_{H^+}>500$\,GeV (at 95\% 
CL) and $m_{H^+}>350$\,GeV (at 99\% CL) found in Ref.~\cite{Gambino:2001ew}.

The function $F(E_0)$ provides us with an alternative way to discuss the 
effects of imposing the cutoff on the photon energy. In contrast to the 
branching ratio, it is independent of several input parameters (e.g., 
$\overline{m}_b(\overline{m}_b)$, $|V_{ts}^* V_{tb}|$, $\tau_B$, 
$\lambda_{1,2}$), and it shows a very weak sensitivity to variations of the 
remaining parameters. We obtain
\begin{equation}
   F(1.8\,{\rm GeV})
   = (92_{\,-10}^{\,+\phantom{1}7}\,[\mbox{pert.}]\pm 1\,[\mbox{pars.}])\% \,.
\end{equation}
This is the first time that this fraction has been computed in a model 
independent way. The result may be compared with the values
$(95.8_{\,-2.9}^{\,+1.3})\%$ and $(95\pm 1)\%$ obtained from two studies of 
shape-function models \cite{Kagan:1998ym,Bigi:2002qq}, in which perturbative 
uncertainties have been ignored. We obtain a significantly smaller central 
value with a much larger uncertainty.

The last quantity we wish to explore is the average photon energy. As mentioned
above, this observable is very sensitive to the interplay of physics at the 
intermediate and low scales. The study of uncertainties due to parameter 
variations exhibits that the prime sensitivity is to the $b$-quark mass, which 
is expected, since $\langle E_\gamma\rangle=m_b/2+\dots$ to leading order. The 
next-important contribution to the error comes from the HQET parameter 
$\lambda_1$. To a very good approximation, we have
\begin{equation}
   \langle E_\gamma\rangle \Big|_{E_0=1.8\,{\rm GeV}}
   = (2.27_{\,-0.07}^{\,+0.05})\,\mbox{GeV}
   + \frac{\delta m_b}{2} - \frac{\delta\lambda_1}{4m_b} \,,
\end{equation}
where the error accounts for the perturbative uncertainty. The quantities 
$\delta m_b$ and $\delta\lambda_1$ parameterize possible deviations of the 
relevant input parameters from their central values $m_b=4.65$\,GeV and 
$\lambda_1=-0.25$\,GeV$^2$. Our prediction is in excellent agreement with the 
Belle result in (\ref{belle}). This finding provides support to the value of 
the $b$-quark mass in the shape-function scheme extracted in 
Ref.~\cite{Bosch:2004th}. We stress, however, that the large perturbative 
uncertainties in the formula for $\langle E_\gamma\rangle$ impose significant 
limitations on the precision with which $m_b$ can be extracted from a 
measurement of the average photon energy. Our estimate above implies a 
perturbative uncertainty of $\delta m_b[{\rm pert.}]={}_{-100}^{+140}$\,MeV. 
This is in addition to twice the experimental error in the measurement of 
$\langle E_\gamma\rangle$, which at present yields 
$\delta m_b[{\rm exp.}]=86$\,MeV.

\section{Conclusions and outlook}

We have performed the first systematic analysis of the inclusive decay 
$B\to X_s\gamma$ in the presence of a photon-energy cut $E_\gamma\ge E_0$, 
where $E_0$ is such that $\Delta=m_b-2E_0$ can be considered large compared to 
$\Lambda_{\rm QCD}$, while still $\Delta\ll m_b$. This is the region of 
interest to experiments at the $B$ factories. The first condition 
($\Delta\gg\Lambda_{\rm QCD}$) ensures that a theoretical treatment without 
shape functions can be applied. However, the second condition 
($\Delta\ll m_b$) means that this treatment is {\em not\/} a conventional 
heavy-quark expansion in powers of $\alpha_s(m_b)$ and 
$\Lambda_{\rm QCD}/m_b$. Instead, we have shown that three distinct 
short-distance scales are relevant in this region. They are the hard scale 
$m_b$, the hard-collinear (or jet) scale $\sqrt{m_b\Delta}$, and the low scale 
$\Delta$. To separate the contributions associated with these scales requires 
a multi-scale operator product expansion (MSOPE). 

Our approach allows us to study analytically the transition from the 
shape-function region, where $\Delta\sim\Lambda_{\rm QCD}$, into the MSOPE
region, where $\Lambda_{\rm QCD}\ll\Delta\ll m_b$, into the region 
$\Delta={\cal O}(m_b)$, where a conventional heavy-quark expansion applies. 
This is a significant improvement over previous work. For instance, it has 
sometimes been argued that exactly where the transition to a conventional 
heavy-quark expansion occurs is an empirical question, which cannot be 
answered theoretically. Our formalism provides a precise, quantitative answer 
to this question. In particular, for $B\to X_s\gamma$ with a realistic cut on 
the photon energy, one is {\em not}\/ in a situation where a short-distance 
expansion at the scale $m_b$ can be justified. The analysis makes it evident 
that the precision that can be achieved in the prediction of the 
$B\to X_s\gamma$ branching ratio is, ultimately, determined by how well 
perturbative and non-perturbative corrections can be controlled at the lowest 
relevant scale $\Delta$, which in practice is of order 1\,GeV. Consequently, 
we estimate much larger theoretical uncertainties than previous authors. These 
uncertainties are dominated by yet unknown higher-order perturbative effects. 
Non-perturbative, hadronic effects at the scale $\Delta$ appear to be small 
and under control. 

This is not the first time in the history of $B\to X_s\gamma$ calculations 
that issues of scale setting have changed the prediction and error estimate 
for the branching ratio (see, e.g., the discussion in 
Ref.~\cite{Gambino:2001ew}). In our case, however, the change in perspective 
about the theory of $B\to X_s\gamma$ decay is more profound, as it imposes 
limitations on the very validity of a short-distance treatment. If the 
short-distance expansion at the scale $\Delta$ fails, then the rate 
{\em cannot\/} be calculated without resource to non-perturbative shape 
functions, which introduces an irreducible amount of model dependence. In 
practice, while $\Delta\approx 1.1$\,GeV (for $E_0\approx 1.8$\,GeV) is 
probably sufficiently large to trust a short-distance analysis, it would be 
unreasonable to expect that yet unknown higher-order effects should be less 
important than in the case of other low-scale applications of QCD.

Obtaining a precise prediction for the $B\to X_s\gamma$ decay rate in the 
Standard Model is an important target of heavy-flavor theory. The present work 
shows that the ongoing effort to calculate the dominant parts of the 
next-to-next-to-leading corrections in the conventional heavy-quark expansion 
is only part of what is needed to achieve this goal. Equally important will be 
to compute the dominant higher-order corrections proportional to 
$\alpha_s^2(\Delta)$ and $\alpha_s^2(\sqrt{m_b\Delta})$, and to perform a 
renormalization-group analysis of the leading kinematic power corrections of 
order $\Delta/m_b$. In fact, our error analysis suggests that these effects 
are potentially more important that the hard matching corrections at the scale 
$m_b$.

\vspace{0.5cm}\noindent
{\em Acknowledgments:\/}
I am grateful to the organizers for the invitation to deliver this talk. It is 
a pleasure to thank Alex Kagan and Bj\"orn Lange for useful discussions. This 
research was supported by the National Science Foundation under Grant 
PHY-0355005.

\end{document}